\documentclass[aps,pra,twocolumn,amsmath,amssymb,nofootinbib,superscriptaddress]{revtex4-1}

\usepackage{times}
\usepackage[pdftex]{graphicx}
\usepackage{dcolumn}
\usepackage{bm}
\usepackage{amsmath}
\usepackage{indentfirst}
\usepackage{float}
\usepackage[colorlinks]{hyperref}
\usepackage[dvipsnames]{xcolor}
\usepackage{xcolor}
\usepackage{subfigure}
\usepackage{algorithm}
\usepackage{algorithmicx}
\usepackage{algpseudocode}
\usepackage{amsmath}
\usepackage{verbatim}
\usepackage{makecell}
\usepackage[normalem]{ulem}

\newcommand{\im}{{\rm i}}
\newcommand{\rate}{W}
\newcommand{\WSC}{{\rm WSC}}
\newcommand{\SSC}{{\rm SSC}}
\newcommand{\nss}{\bar{n}_{{\rm st}}}
\newcommand{\rhoop}{\hat{\rho}}
\newcommand{\Hop}{\hat{H}}

\newcommand{\Dop}{\mathcal{D}}
\newcommand{\aop}{\hat{a}}
\newcommand{\adop}{\hat{a}^{\dagger}}
\newcommand{\xop}{\hat{x}}
\newcommand{\LD}{{\rm LD}}
\newcommand{\Xop}{\hat{X}}
\newcommand{\Hc}{{\rm H.c.}}
\newcommand{\sps}{\mathcal{M}}
\newcommand{\nave}{\bar{n}}

\newcommand{\hnu}{Key Laboratory of Low-Dimensional Quantum Structures and Quantum Control of Ministry of Education, Department of Physics and Synergetic Innovation Center for Quantum Effects and Applications, Hunan Normal University, Changsha 410081, China
}

\begin{document}

\title{Sideband Cooling of a Trapped Ion in Strong Sideband Coupling Regime}

\author{Shuo Zhang}
\affiliation{Henan Key Laboratory of Quantum Information and Cryptography, Zhengzhou,
Henan 450000, China}

\author{Zhuo-Peng Huang}
\affiliation{Henan Key Laboratory of Quantum Information and Cryptography, Zhengzhou,
Henan 450000, China}

\author{Tian-Ci Tian}
\affiliation{Henan Key Laboratory of Quantum Information and Cryptography, Zhengzhou,
Henan 450000, China}

\author{Zheng-Yang Wu}
\affiliation{Henan Key Laboratory of Quantum Information and Cryptography, Zhengzhou,
Henan 450000, China}

\author{Jian-Qi Zhang}
\affiliation{State Key Laboratory
of Magnetic Resonance and Atomic and Molecular Physics, 
Wuhan Institute of Physics and Mathematics, Innovation Academy
of Precision Measurement Science and Technology, Chinese
Academy of Sciences, Wuhan 430071, China}

\author{Wan-Su Bao}
\affiliation{Henan Key Laboratory of Quantum Information and Cryptography, Zhengzhou,
Henan 450000, China}

\author{Chu Guo}
\email{guochu604b@gmail.com}
\affiliation{Henan Key Laboratory of Quantum Information and Cryptography, Zhengzhou,
Henan 450000, China}
\affiliation{\hnu}



\begin{abstract}
Conventional theoretical studies on the ground-state laser cooling of
a trapped ion have mostly focused on the weak sideband coupling (WSC) regime,
where the cooling rate is inverse proportional to the linewidth
of the excited state. In a recent work~[New J. Phys. 23, 023018 (2021)], we proposed a theoretical framework
to study the ground state cooling of a trapped ion in the strong sideband coupling (SSC)
regime, under the assumption of a vanishing carrier transition. Here we extend this analysis to more general situations with nonvanishing carrier transitions, where we show that by properly tuning the coupling lasers a cooling rate proportional to the linewidth can be achieved.
Our theoretical predictions closely agree with the corresponding exact solutions in the SSC regime, which provide an important theoretical guidance for  sideband cooling experiments.
\end{abstract}

\maketitle

\section{Introduction}

Trapped ions system has become one of the most promising candidates
for quantum computing~\cite{CiracZoller1995,haffner2008quantum,bruzewicz2019trapped}.
Compared to other systems, a string of ions trapped in a linear Paul
trap enjoys several advantages such as a very long decoherence time~\cite{bermudez2017assessing,wang2017single}, high-fidelity gate
operations and readout~\cite{ballance2016high,gaebler2016high,srinivas2021high,myerson2008high,crain2019high},
as well as fully-connected architecture~\cite{debnath2016demonstration,pino2021demonstration,niroula_constrained_2022}.
A pre-request for high-fidelity multi-qubit gate operations
in trapped ions system is to cool down the motional degrees of freedom
(phonons) of the ions close to their ground states.

Sideband cooling has been the method of choice to cool down a trapped ion
into its motional ground state for a long time due to its simplicity and excellent performance in practice~\cite{DiedrichWineland1989,MonroeWineland1995,peik1999sideband,roos1999quantum,deslauriers2004zero,hemmerling2011single,seck2016raman,goodwin2016resolved,che2017efficient,chen2017sympathetic,stutter2018sideband,hrmo2019sideband,joshi2019population,chen2020efficient,wu2022continuous}.
In its ideal settings, the standard sideband cooling only requires to couple the trapped ion to a single running wave laser, where the laser is able to excite the ion from a (meta-)stable ground state to an unstable excited state, and that the ion is pre-cooled into the Lamb-Dicke regime characterized by a small dimensionless Lamb-Dicke parameter $\eta$ ($\eta\ll 1$ means that the motion of the ion is negligible compared to the wave length of the laser, which could be achieved using some pre-cooling methods such as Doppler cooling~\cite{DiedrichWineland1989} and polarization gradient cooling~\cite{birkl1994polarization,ejtemaee20173d,joshi2020polarization,li2022robust}).
The dissipative dynamics of the trapped ion in the standard sideband cooling is determined by the interplay of the four system parameters: the coupling strength $\Omega$ and the detuning $\Delta$ of the running wave laser, the linewidth $\gamma$ of the unstable excited state and the phonon energy $\nu$ (taking $\hbar=1$). In the resolved sideband limit, namely $\eta \Omega, \gamma \ll \nu$, the dynamics is dominated by three transition lines: the carrier transition (with strength $\Omega$) that the ion absorbs one photon
without any change of the motional state, the weaker red (blue) sideband transitions (with strength $\eta\Omega$) that the ion absorbs one photon and at the same time the motional state decreases (increases) by one phonon.

In the weak sideband coupling (WSC) limit, where the sideband coupling strength is much smaller compared to the linewidth of the excited state, namely $\eta\Omega \ll \gamma$, the cooling mechanism has been well understood~\cite{neuhauser1978optical,wineland1979laser,stenholm1986semiclassical,CiracZoller1992}: by tuning the laser to red sideband resonance with the condition
\begin{align}\label{eq:cond_wsc}
 \Delta=-\nu, 
\end{align}
the internal state of the ion will mostly stay in the ground state and can be adiabatically eliminated from the motional state during the cooling process, while the dynamics of the motional state can be effectively described as an exponential decay of the average phonon occupation $\nave$ with a rate
\begin{align}\label{eq:rate_wsc}
\rate^{\WSC} \propto \frac{\eta^2\Omega^2}{\gamma},
\end{align}
and the final steady state average phonon occupation
\begin{align}\label{eq:nss_wsc}
\nss^{\WSC} \approx (\alpha + \frac{1}{4})\left(\frac{\gamma}{2\nu}\right)^2,
\end{align}
where the geometry factor $\alpha = \frac{2}{5}$ for dipole transition, and only the zeroth order term of $\eta$ is kept which is due to the blue sideband heating transition. 
Subsequent improvements over the original sideband cooling scheme mainly focus on suppressing the heating effects induced by the carrier
and the blue sideband transitions by preparing the ground state to be the dark state of those transitions (but still in the WSC regime). Outstanding examples in this direction include the standing wave sideband cooling~\cite{CiracZoller1992} and the electromagnetically induced transparency
(EIT) assisted cooling where the carrier transitions are eliminated~\cite{MorigiKeitel2000,RoosBlatt2000,Morigi2003,LinWineland2013,KampschulteMorigi2014,Lechner2016Roos,ScharnhorstSchmidt2018,JordanBollinger2019,FengMonroe2020,ZhangGuo2021,QiaoKim2021,zhang2022parallel}, as well as more sophisticated schemes which also eliminate the blue sideband transitions~\cite{EversKeitel2004,RetzkerPlenio2007,CerrilloPlenio2010,AlbrechtPlenio2011,ZhangChen2012,ZhangChen2014,LuGuo2015,CerrilloPlenio2018,wang2022superior,wang2022enhanced}. 



In comparison, the cooling mechanism in the strong sideband coupling (SSC) regime with $\eta\Omega \geq \gamma$ remains largely unexplored for the standard sideband cooling. Existing cooling schemes have mostly considered the WSC regime and focused on minimizing $\nss$, the price to pay is a very low cooling rate which is inverse proportional to $\gamma$ as in Eq.(\ref{eq:rate_wsc}). 
However, from an experimental point of view, a large cooling rate is at least of the same importance as a low steady state average phonon occupation.
In our previous work, we have proposed a theoretical
framework to understand the ground state cooling of a trapped ion in the SSC regime under the assumption of vanishing carrier transitions~\cite{ZhangGuo2021}, where we show that a fast cooling rate proportional to $\gamma$ can be achieved under a modified red sideband resonance condition in the dressed state picture of both the internal and motional degrees of freedom. However, the derivations there can not be directly extended to cooling
schemes with nonvanishing carrier transitions such as the standard sideband cooling.

In this work, we develop a theoretical framework to understand the cooling mechanism in the SSC regime and in presence of a nonvanishing carrier transition, where we have used a new dressed state picture which could incorporate both the resonant red sideband transition and the off-resonant carrier transition. Similar to Ref.~\cite{ZhangGuo2021}, we predict a cooling rate which is proportional to the linewidth of the excited state, but with a significant difference that the cooling rate will also be dependent on the laser coupling strength. In the meantime, we show that the major heating effect in the SSC regime is the carrier excitation in the dressed state picture, strikingly different from the WSC regime. 
The paper is organized as follows. In Sec.~\ref{sec:model} we introduce the standard sideband cooling model as well as our analytic solutions of it in the SSC regime. Then we verify our theoretical predictions against the numerical solutions of the exact quantum master equation in Sec.~\ref{sec:results}. We conclude in Sec.~\ref{sec:summary}.



\section{Sideband cooling in the SSC regime}\label{sec:model}

\begin{figure*}[htbp]
\begin{centering}
\includegraphics[width=2\columnwidth]{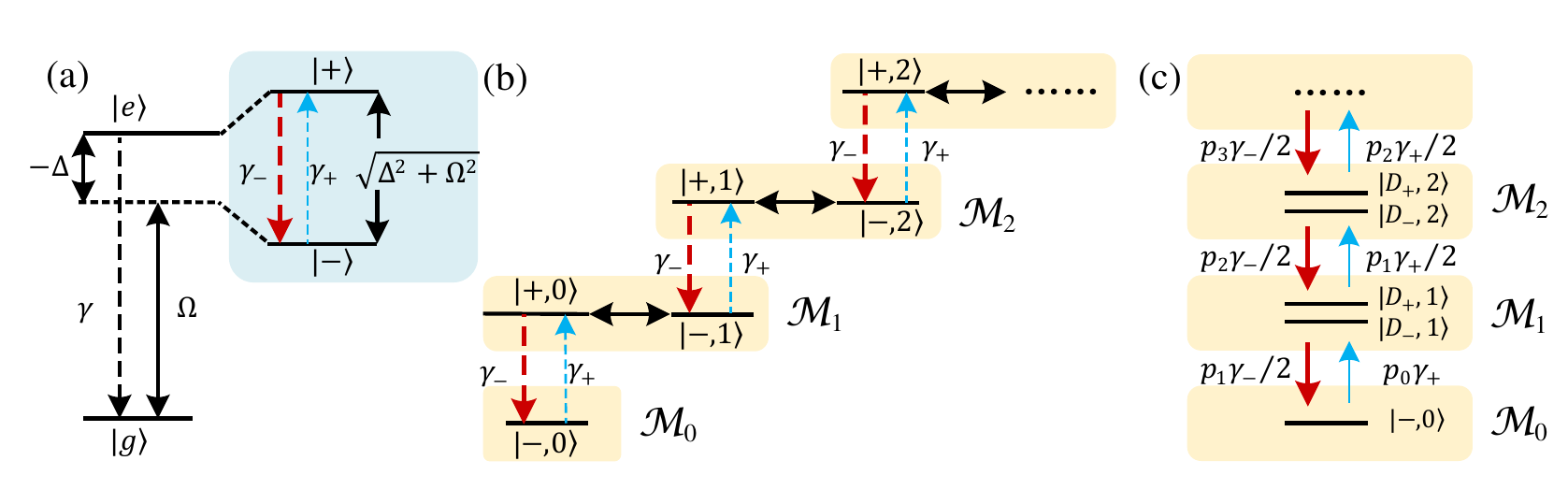} 
\par\end{centering}
\caption{(a) Level structure for the standard sideband cooling. The system consists
of an unstable excited state $\left|e\right\rangle $ with linewidth $\gamma$ and a ground
state $\left|g\right\rangle $, which are coupled by a running wave laser field with detuning $\Delta$ and strength $\Omega$. The internal Hamiltonian
can be diagonalized with two dressed states $\left|+\right\rangle $ and $\left|-\right\rangle$, with an effective dissipation $\left|+\right\rangle \protect\rightarrow\left|-\right\rangle $ with rate $\gamma_-$ and an effective dissipation $\left|-\right\rangle \protect\rightarrow\left|+\right\rangle $ with rate $\gamma_+$ respectively. 
(b) The cooling dynamics in the dressed state basis $\vert \pm\rangle$. The whole Hilbert space can be divided into subspaces $\{\sps_0, \sps_1, \dots\}$ with $\sps_{0}=\left\{ \left|-,0\right\rangle \right\} $ and $\sps_{n}=\left\{ \left|+,n-1\right\rangle ,\left|-,n\right\rangle \right\} $. And the
optimal cooling condition is chosen such that the transition $\left|-,n\right\rangle \protect\leftrightarrow\left|+,n-1\right\rangle $
is resonant. 
(c) In SSC regime, one can further diagonalize each subspace $\sps_n$ with $n\geq 1$ to get the dressed states $\left|D_{+},n\right\rangle $
and $\left|D_{-},n\right\rangle $. Then cooling dynamics can be understood as a cascade between the subspaces from up down.
}
\label{fig:fig1} 
\end{figure*}

A minimal theoretical model for the standard sideband cooling contains a trapped ion with two internal states: a (meta-)stable ground state denoted as $\vert g\rangle$ and an unstable excited state denoted as $\vert e\rangle$, together with a running wave laser that drives the transition $\left|g\right\rangle \leftrightarrow\left|e\right\rangle $. The energy difference between $\vert g\rangle$ and $\vert e\rangle$ is denoted as $\omega_0$. The strength and the frequency of the laser are denoted as $\Omega$ and $\omega_L$ respectively. The model is shown in Fig.~\ref{fig:fig1}(a).
The equation of motion is described by the Lindblad master equation~\cite{GoriniSudarshan1976,Lindblad1976}
\begin{align}\label{eq:lindblad}
\frac{d\rhoop}{dt} = -\im [\Hop, \rhoop] + \Dop(\rhoop),
\end{align}
with the Hamiltonian 
\begin{align}
\Hop =& -\Delta \vert e\rangle\langle e\vert + \nu \adop\aop \nonumber \\ 
&+ \frac{\Omega}{2}(\vert e\rangle\langle g\vert e^{-\im k_L\xop} + \vert g\rangle\langle e\vert e^{\im k_L\xop} ), 
\end{align}
where $\Delta=\omega_L - \omega_0$ is the detuning, $\nu$ is the trap frequency (phonon energy), $\adop$ and $\aop$ are the creation and annihilation operators for the motional state, $k_L = \omega_L / c$ ($c$ is the speed of light) and $\xop=\sqrt{\frac{1}{2m\nu}}(\adop+\aop)$ ($m$ is the mass of the ion). The dissipation can be described as
\begin{align}
\Dop(\rhoop) =& \frac{\gamma}{2}\int_{-1}^1 d(\cos(\theta)) \left( \frac{3}{4}(1 + \cos^2(\theta)) \right) \vert g\rangle\langle e\vert \times \nonumber \\
& e^{\im k_L\xop\cos(\theta)}\rhoop e^{-\im k_L\xop\cos(\theta)}\vert e\rangle\langle g\vert - \frac{\gamma}{2}\{\vert e\rangle\langle e\vert, \rhoop\}.
\end{align}
In Lamb-Dicke regime, we expand $\Hop$ to the first order of $\eta$ and get
\begin{align}
\Hop^{\LD} = \Hop_0 + \Hop_1,
\end{align}
where the zeroth order term is
\begin{align}\label{eq:H0}
\Hop_0 = \nu \adop \aop-\Delta\left|e\right\rangle \left\langle e\right|+\frac{\Omega}{2}\left(\left|e\right\rangle \left\langle g\right|+\left|g\right\rangle \left\langle e\right|\right),
\end{align}
and the first order term is
\begin{align}\label{eq:H1}
\Hop_{1} = \eta\frac{\Omega}{2}\left(-\im\left|e\right\rangle \left\langle g\right|+\im\left|g\right\rangle \left\langle e\right|\right)\left(\adop+\aop\right).
\end{align}
The dissipator $\Dop$ is kept to the zeroth order of $\eta$, which is
\begin{align}\label{eq:D0}
\Dop_0(\rhoop) = \gamma\left( \vert g\rangle\langle e\vert \rhoop \vert e\rangle\langle g\vert - \frac{1}{2}\{\vert e\rangle\langle e\vert, \rhoop \} \right).
\end{align}

Now we first diagonalize the internal states of $\Hop_0$ from Eq.(\ref{eq:H0}) and get
\begin{align}\label{eq:H0dressed}
\Hop_0=\nu \adop\aop+\omega_{+}\left|+\right\rangle \left\langle +\right|+\omega_{-}\left|-\right\rangle \left\langle -\right|,
\end{align}
where the two dressed states
\begin{align}
\left|+\right\rangle  & =  \sin\phi\left|e\right\rangle +\cos\phi\left|g\right\rangle; \label{eq:plus} \\
\left|-\right\rangle  & =  \cos\phi\left|e\right\rangle -\sin\phi\left|g\right\rangle \label{eq:minus}
\end{align}
correspond to the dressed energies $\omega_{\pm}=\frac{1}{2}\left(-\Delta\pm\sqrt{\Delta^{2}+\Omega^{2}}\right)$ with $\phi=\arcsin\left(\sqrt{\frac{1}{2}\left(1-\frac{\Delta}{\sqrt{\Delta^{2}+\Omega^{2}}}\right)}\right)$.
Then we can rewrite $\Hop_{1}$ of Eq.(\ref{eq:H1}) in dressed state basis $\vert \pm\rangle$ as
\begin{align} \label{eq:H1dressed}
\Hop_{1} = & \eta\frac{\Omega}{2}\left(\im\left|+\right\rangle \left\langle -\right|-\im\left|-\right\rangle \left\langle +\right|\right)\left(\adop+\aop\right).
\end{align}
From Eqs.(\ref{eq:H0dressed}, \ref{eq:H1dressed}), we can see that the Hamiltonian in the dressed
state basis $\vert \pm\rangle$ is equivalent to the Hamiltonian of a standing wave sideband cooling model in the bare basis $\{\vert g\rangle, \vert e\rangle\}$, using the mappings
$\left|-\right\rangle \leftrightarrow\left|g\right\rangle $,
$\im\left|+\right\rangle \leftrightarrow\left|e\right\rangle $, and
$\omega_{-}-\omega_{+}\leftrightarrow\Delta$.
The dissipation $\Dop_0$ from Eq.(\ref{eq:D0}) can be written in dressed state basis $\vert \pm\rangle$  as
\begin{align}\label{eq:D0dressed}
\Dop_0(\rhoop) = & \frac{\gamma_{-}}{2} \Dop_{\vert -\rangle \langle + \vert}(\rhoop) + \frac{\gamma_{+}}{2} \Dop_{\vert +\rangle \langle - \vert}(\rhoop) \nonumber \\ 
&+ \frac{\gamma_{\phi}}{2} \Dop_{\vert +\rangle \langle + \vert - \vert -\rangle \langle - \vert}(\rhoop),
\end{align}
where we have used $\Dop_{\Xop}(\rhoop) = 2\Xop\rhoop \Xop^{\dagger} - \{\Xop^{\dagger}\Xop, \rhoop\} $ for an operator $\Xop$. 
The dissipation rates in Eq.(\ref{eq:D0dressed}) are given by $\gamma_{\pm}=\left(\frac{1\mp\beta}{2}\right)^{2} \gamma$
and $\gamma_{\phi}=\frac{1-\beta^{2}}{4} \gamma$, with the parameter $\beta$ defined as 
\begin{align}\label{eq:beta}
\beta=\frac{\sqrt{\nu^{2}-\Omega^{2}}}{\nu}.
\end{align}
The two terms on the first line of the right hand side of Eq.(\ref{eq:D0dressed}) correspond to the effective dissipations from $\vert +\rangle$ to $\vert -\rangle$ and from $\vert -\rangle$ to $\vert +\rangle$, which induce cooling and heating effects respectively as shown in Fig.~\ref{fig:fig1}(a,b), 
while the term on the second line of Eq.(\ref{eq:D0dressed}) is a pure dephasing term. In the WSC regime, we have $\beta\approx 1$, and thus $\gamma_{-} \approx \gamma$ and $\gamma_{+}\approx \gamma_{\phi}\approx 0$.


Due to the correspondence between the Hamiltonian for the running wave sideband cooling in the dressed state basis $\vert \pm\rangle$ and the Hamiltonian for the standing wave sideband cooling in the bare basis, the optimal cooling condition in the former case can be naturally obtained by following the derivation in the latter case~\cite{CiracZoller1992}.
In the next we go to the interaction picture for $\Hop^{\LD}$, where the density operator takes the form 
\begin{align}
\rhoop^I=e^{\im\Hop_{0}t}\rhoop e^{-\im\Hop_{0}t},
\end{align}
and the Hamiltonian takes the form 
\begin{align}\label{eq:Hint}
\Hop^{I} = & e^{-\im\Hop_{0}t}\Hop_{1}e^{\im\Hop_{0}t}\nonumber \\
 = & \eta\frac{\Omega}{2}\left(-\im\left|-\right\rangle \left\langle +\right|e^{\im\left(\omega_{-}-\omega_{+}\right)t}+\Hc\right)\left(\adop e^{-\im\nu t}+\Hc\right).
\end{align}
Then similar to the standing wave sideband cooling, the optimal cooling condition can be chosen as $\omega_{-}-\omega_{+}=-\nu$, that is, 
\begin{equation}\label{eq:cond_ssc}
\Delta=-\sqrt{\nu^{2}-\Omega^{2}},
\end{equation}
such that the red sideband transition $\left|-,n\right\rangle \leftrightarrow\left|+,n-1\right\rangle $$\left(n\geq1\right)$ ($\vert n\rangle$ denotes the motional state with $n$ phonons) in the dressed state basis becomes resonant. 
Here we note that the cooling condition
in Eq.(\ref{eq:cond_ssc}) has also been identified in Refs.~\cite{marzoli1994laser,li2021fast}.
From Eqs.(\ref{eq:beta}, \ref{eq:cond_ssc}) we can also see that $\beta = -\frac{\Delta}{\nu}$.
Under the new red sideband resonant condition in Eq.(\ref{eq:cond_ssc}), we further neglect the blue sideband transition term in Eq.(\ref{eq:Hint}) by making the rotating wave approximation, and obtain the time-independent red sideband resonant Hamiltonian $\Hop_r^I$ as
\begin{equation}\label{eq:Hr}
\Hop_r^I=\frac{\eta\Omega}{2}\left(\im\left|+\right\rangle \left\langle -\right|\aop-\im\left|-\right\rangle \left\langle +\right|\adop\right).
\end{equation}
One reason that we can make this approximation is that in the SSC regime the blue sideband transition is no longer the dominate heating term, as will be clear later.
The dissipation $\Dop_0$ remains the same in the interaction picture, namely $\Dop^I_0 = \Dop_0$ and we will still use $\Dop_0$ afterwards. As a result the exact master equation in Eq.(\ref{eq:lindblad}) can be simplified to 
\begin{align}\label{eq:lindblad2}
\frac{d}{dt}\rhoop^{I}\left(t\right)=\left[\Hop_{r}^I,\rhoop^{I}\left(t\right)\right]+\mathcal{D}_{0}\left(\rhoop^{I}\left(t\right)\right).
\end{align}

Now we proceed to derive the analytical expressions for the cooling rate and the steady state average phonon occupation in the SSC regime by solving Eq.(\ref{eq:lindblad2}).
In the dressed state basis $\vert \pm\rangle$, the carrier transition vanishes as can be seen from Eq.(\ref{eq:Hr}), which is the same as the Hamiltonian considered in Ref.~\cite{ZhangGuo2021}, while the dissipation in Eq.(\ref{eq:D0dressed}) contains two additional terms
$\Dop_{\vert +\rangle\langle -\vert}$ and $\Dop_{\vert +\rangle\langle +\vert - \vert -\rangle\langle-\vert}$ compared to Ref.~\cite{ZhangGuo2021}.
Therefore similar to Ref.~\cite{ZhangGuo2021}, in the SSC regime 
the resonant transition $\left|-,n\right\rangle \leftrightarrow\left|+,n-1\right\rangle $
oscillates faster than the dissipative terms, therefore the two states $|-,n\rangle$ and $\left|+,n-1\right\rangle$ should be treated together as a whole.
Concretely, we divide the states of the whole system into
subspaces labeled as $\{\sps_0, \sps_1, \dots\}$, where $\sps_{0}=\left\{ \left|-,0\right\rangle \right\} $ and $\sps_{n}=\left\{ \left|+,n-1\right\rangle ,\left|-,n\right\rangle \right\} $ for $n\geq 1$,
and we can further diagonalize $\Hop^I_r$ as
\begin{align}\label{eq:Hr2}
\Hop^I_r=\frac{\eta\Omega}{2} \sum_{n=1}^{\infty}\sqrt{n}\left(\left|D_{+},n\right\rangle \left\langle D_{+},n\right|-\left|D_{-},n\right\rangle \left\langle D_{-},n\right|\right),
\end{align}
where the states
\begin{align}
\left|D_{+},n\right\rangle  = & \frac{1}{\sqrt{2}}\left(\left|-,n\right\rangle +i\left|+,n-1\right\rangle \right);  \\
\left|D_{-},n\right\rangle  = & \frac{1}{\sqrt{2}}\left(\left|-,n\right\rangle -i\left|+,n-1\right\rangle \right)
\end{align}
are equal superpositions of the states $\left|-,n\right\rangle $ and $\left|+,n-1\right\rangle $.
Eq.(\ref{eq:lindblad2}) is still too complicated for an analytic solution. Similar to Ref.~\cite{ZhangGuo2021}, we neglect the off-diagonal terms of $\rhoop^I$ and use the following ansatz for $\rhoop^I$:
\begin{align}\label{eq:rhoI}
\rhoop^I\left(t\right) = & p_{0}\left(t\right)\left|-,0\right\rangle \left\langle -,0\right| + \nonumber \\
 & \sum_{n=1}^{\infty}\left(p_{n,+}\left(t\right)\left|D_{+},n\right\rangle \left\langle D_{+},n\right|\right.\nonumber \\
 & \left.+p_{n,-}\left(t\right)\left|D_{-},n\right\rangle \left\langle D_{-},n\right|\right).
\end{align}
Substitute Eq.(\ref{eq:rhoI}) into Eq.(\ref{eq:lindblad2}), we get the equations of motion for the diagonal coefficients $p_{n,\pm}(t)$ as
\begin{subequations}\label{eq:rate0}
\begin{align}
\frac{d}{dt}p_{0} = & -\gamma_{+}p_{0}+\frac{\gamma_{-}}{2}p_{1,+}+\frac{\gamma_{-}}{2}p_{1,-}; \\
\frac{d}{dt}p_{1,+} = & \frac{\gamma_{+}}{2}p_{0}-\left(\frac{\gamma_{+}}{2}+\frac{\gamma_{-}}{2}+\gamma_{\phi}\right)p_{1,+} \nonumber \\
 & -\gamma_{\phi}p_{1,-}+\frac{\gamma_{-}}{4}p_{2,+}+\frac{\gamma_{-}}{4}p_{2,-}; \label{eq:rate0b} \\
\frac{d}{dt}p_{1,-} = & \frac{\gamma_{+}}{2}p_{0}-\left(\frac{\gamma_{+}}{2}+\frac{\gamma_{-}}{2}+\gamma_{\phi}\right)p_{1,-} \nonumber \\
 & -\gamma_{\phi}p_{1,+}+\frac{\gamma_{-}}{4}p_{2,+}+\frac{\gamma_{-}}{4}p_{2,-}; \label{eq:rate0c} \\
\frac{d}{dt}p_{n,+} = & \frac{\gamma_{+}}{4}p_{n-1,+}+\frac{\gamma_{+}}{4}p_{n-1,-}- \nonumber \\ 
&\left(\frac{\gamma_{+}}{2}+\frac{\gamma_{-}}{2}+\gamma_{\phi}\right)p_{n,+} -\gamma_{\phi}p_{n,-}+ \nonumber \\ 
&\frac{\gamma_{-}}{4}p_{n\text{+1,+}}+\frac{\gamma_{-}}{4}p_{n+1,-} \quad (n\geq 2); \label{eq:rate0d} \\
\frac{d}{dt}p_{n,-} = & \frac{\gamma_{+}}{4}p_{n-1,+}+\frac{\gamma_{+}}{4}p_{n-1,-} \nonumber \\ 
&-\left(\frac{\gamma_{+}}{2}+\frac{\gamma_{-}}{2}+\gamma_{\phi}\right)p_{n,-} -\gamma_{\phi}p_{n,+}+ \nonumber \\ 
&\frac{\gamma_{-}}{4}p_{n+1,+}+\frac{\gamma_{-}}{4}p_{n+1,-} \quad (n\geq 2) . \label{eq:rate0e}
\end{align}
\end{subequations}
Now we define $p_{n}=p_{n,+}+p_{n,-}$ for $n>0$, which is the probability of the density operator $\rhoop^I$ in the subspace $\sps_n$. Then by summing Eq.(\ref{eq:rate0b}) and Eq.(\ref{eq:rate0c}), and summing Eq.(\ref{eq:rate0d}) and Eq.(\ref{eq:rate0e}), we obtain the rate equations for $p_{n}$ as
\begin{subequations}\label{eq:rate}
\begin{align}
\frac{d}{dt}p_{0} = & -\gamma_{+}p_{0}+\frac{\gamma_{-}}{2}p_{1}; \label{eq:ratea} \\
\frac{d}{dt}p_{1} = & \gamma_{+}p_{0}-\left(\frac{\gamma_{+}+\gamma_{-}}{2}\right)p_{1}+\frac{\gamma_{-}}{2}p_{2}; \label{eq:rateb}  \\
\frac{d}{dt}p_{n} = & \frac{\gamma_{+}}{2}p_{n-1}-\left(\frac{\gamma_{+}+\gamma_{-}}{2}\right)p_{n} \nonumber \\
&+\frac{\gamma_{-}}{2}p_{n+1} \quad (n > 1). \label{eq:ratec}
\end{align}
\end{subequations}
The schematic illustration of Eqs.(\ref{eq:rate}) is shown in Fig.\ref{fig:fig1}(c). Interestingly, although we have neglected the dominant heating
processes in the WSC regime, namely the blue sideband transition and the recoil from the spontaneous decay as can be seen in Eq.(\ref{eq:Hr}) and Eq.(\ref{eq:D0}), there still
exists heating transitions due to the existence of the term $\Dop_{\vert +\rangle\langle -\vert}$ in Eq.(\ref{eq:D0dressed}).


The steady state solutions of the rate equations as in Eqs.(\ref{eq:rate}) can be straightforwardly obtained by setting $\frac{d}{dt}p_{n}=0$ for $n\geq 0$, which are 
\begin{align}
p_{0,\textrm{st}} = & \frac{1-\frac{\gamma_{+}}{\gamma_{-}}}{1+\frac{\gamma_{+}}{\gamma_{-}}}; \\
p_{n,\textrm{st}} = & 2p_{0,\textrm{st}}\left(\frac{\gamma_{+}}{\gamma_{-}}\right)^{n} \quad (n\geq1),
\end{align}
with which the steady state average phonon occupation can be computed as
\begin{align}\label{eq:nss}
\nss^{\SSC} = & \sum_{n=1}^{\infty}p_{n,\text{st}}\left(n-\frac{1}{2}\right)=\frac{\gamma_{+}}{\gamma_{-}-\gamma_{+}}\nonumber \\
 = & \frac{1}{4}\left(\beta+\frac{1}{\beta}\right)-\frac{1}{2}.
\end{align}
Calculating the cooling rate is much more involved, since the first terms on the right hand side of Eqs.(\ref{eq:rateb}, \ref{eq:ratec}) differ by a factor $1/2$. To estimate the cooling rate analytically, we scale the first term on the right hand side of Eq.(\ref{eq:rateb}) by $1/2$ to force it into the same form of Eq.(\ref{eq:ratec}), namely
\begin{subequations}\label{eq:rate_s}
\begin{align}
\frac{d}{dt}p_{0} = & -\frac{\gamma_{+}}{2}p_{0}+\frac{\gamma_{-}}{2}p_{1};  \\
\frac{d}{dt}p_{n} = & \frac{\gamma_{+}}{2}p_{n-1}-\left(\frac{\gamma_{+}+\gamma_{-}}{2}\right)p_{n} \nonumber \\ 
&+\frac{\gamma_{-}}{2}p_{n+1} \quad (n\geq1) .
\end{align}
\end{subequations}
Physically, this approximation means that the heating from $\sps_0$ to $\sps_1$ will be underestimated, therefore the cooling rate obtained from Eqs.(\ref{eq:rate_s}) will be slightly larger than Eqs.(\ref{eq:rate}).
The solution of Eqs.(\ref{eq:rate_s}) will in general be dependent on the initial state, which is in comparison with the WSC limit. Here we will consider the initial motional state in a thermal distribution with an initial average phonon occupation denoted as $n_0$ (the initial internal state is $\vert g\rangle$):
\begin{align}
p_{n}(0) =\left(\frac{1}{1+n_{0}}\right)\left(\frac{n_{0}}{1+n_{0}}\right)^{n}.
\end{align}
Eqs.(\ref{eq:rate_s}) can then be exactly solved using the following ansatz
\begin{align}\label{eq:pnt}
p_{n}\left(t\right)=\left(p_{n}(0)-p_{n,\textrm{st}}\right)e^{-\rate^{\SSC} t}+p_{n,\textrm{st}},
\end{align}
with the cooling rate $\rate^{\SSC}$ to be determined. Substituting the ansatz in Eq (\ref{eq:pnt}) into Eqs.(\ref{eq:rate_s}), we obtain the time evolution for average phonon occupation $\bar{n}(t) =\textrm{tr}\left(\adop\aop\rhoop^I\left(t\right)\right)$ as
\begin{align}\label{eq:nt}
\bar{n}(t) =n_{0}\left(1-\frac{1}{2\left(1+n_{0}\right)}\right)e^{-\rate^{\SSC}t}+\nss^{\SSC},
\end{align}
with the cooling rate 
\begin{align}\label{eq:rate_ssc}
\rate^{\SSC} = & \frac{1}{1+n_{0}}\frac{\gamma_{-}}{2}-\frac{1}{n_{0}}\frac{\gamma_{+}}{2}.
\end{align}
Assuming that the ion is initially at
the Doppler cooling limit with $n_{0}\gg1$, we have
\begin{align}\label{eq:rate_ssc2}
\rate^{\SSC} \approx \frac{\gamma}{2n_{0}}\beta  .
\end{align}
Now we note that the cooling rate in Eq.(\ref{eq:rate_ssc2}) is similar to the one obtained
for the standing wave sideband cooling in the SSC regime (which is $\frac{\gamma}{2\left(n_{0}+1\right)}$~\cite{ZhangGuo2021}).
Both of them are proportional to the (effective) linewidth $\gamma$ and are inverse proportional to $n_{0}$. However, the cooling rate in Eq.(\ref{eq:rate_ssc2}) is also dependent on the laser coupling strength $\Omega$ through $\beta$, concretely $\rate^{\SSC}$ decreases when increasing $\Omega$. This is a special feature for nonvanishing carrier transitions, which also means that the cooling rate in case of nonvanishing carrier transitions will generally be smaller than that for the case of vanishing carrier transitions.


\section{Numerical Results}\label{sec:results}

\begin{figure}[htbp]
\begin{centering}
\includegraphics[width=1\columnwidth]{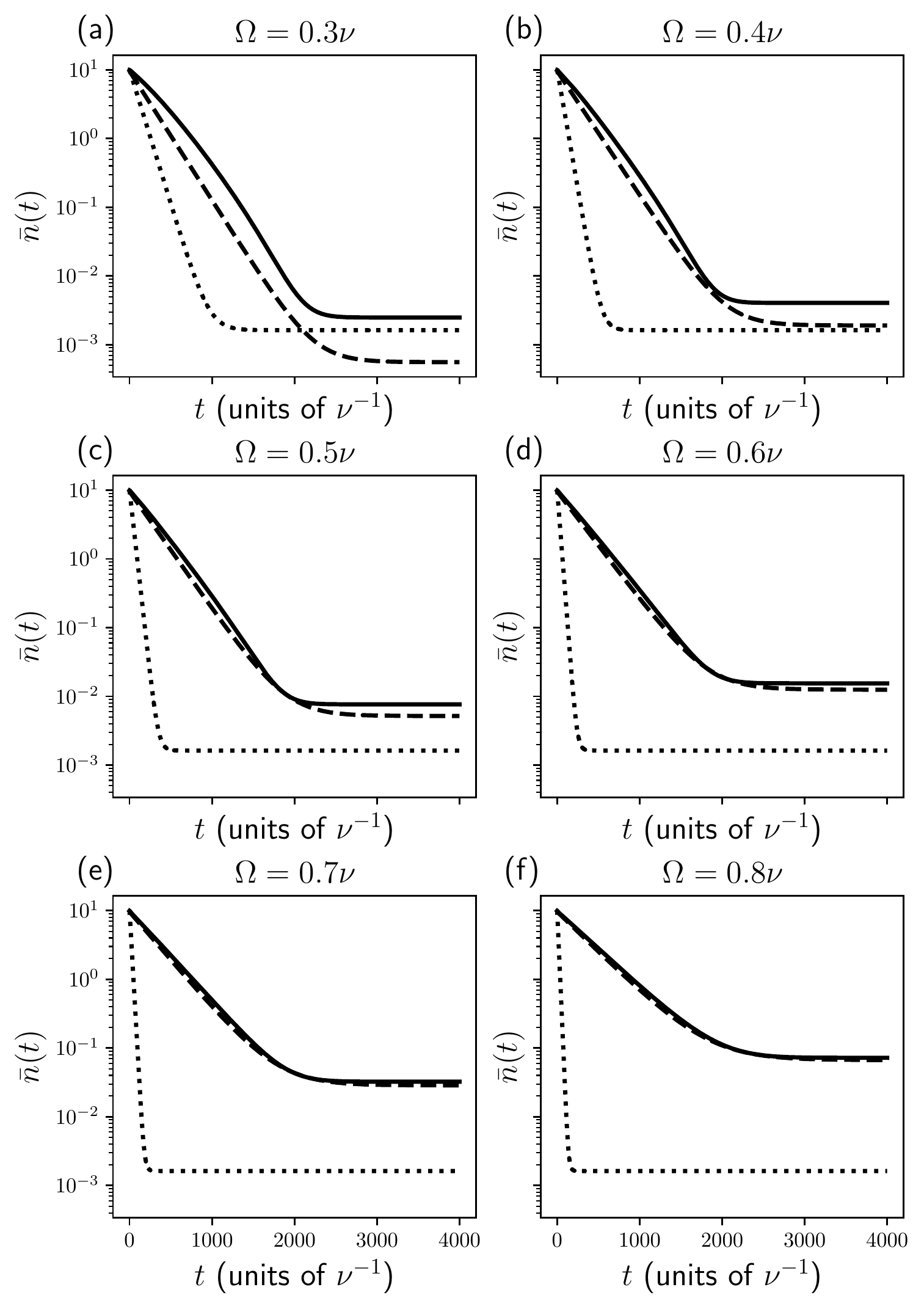} 
\par\end{centering}
\caption{Average phonon occupation $\nave(t) $
as functions of time $t$ for different coupling strength $\Omega$. The solid lines are numerical solutions of the exact master equation in Eq.(\ref{eq:lindblad}), the dashed lines are the analytical predictions using our SSC theory in Eq.(\ref{eq:nt}), and the dotted lines are predictions using the WSC theory in Eqs.(\ref{eq:rate_wsc}, \ref{eq:nss_wsc}). The other parameters used are  $\eta=0.1$, $\gamma=0.1\nu$ and $n_{0}=10$. We have used a truncation of $70$ for the Fock space of the motional state in our numerical simulations throughout this work.
}
\label{fig:fig2} 
\end{figure}

To validate our analytical predictions for the cooling rate and the steady state average phonon occupation, we first compare the approximate cooling dynamics predicted by Eq.(\ref{eq:nt}) with the exact dynamics from Eq.(\ref{eq:lindblad}) for different laser coupling strength $\Omega$, which is shown in Fig~.\ref{fig:fig2}.
We can see that in all the cases considered, our analytical predictions are much more precise compared to the predictions using Eqs.(\ref{eq:rate_wsc}, \ref{eq:nss_wsc}) which are derived in the WSC regime. Moreover, our analytical predictions become more precise as $\Omega$ becomes larger, which is as expected since for larger $\Omega$ we are better in the SSC regime (indicated by the ratio $\eta\Omega/\gamma$).



\begin{figure}[htbp]
\begin{centering}
\includegraphics[width=1\columnwidth]{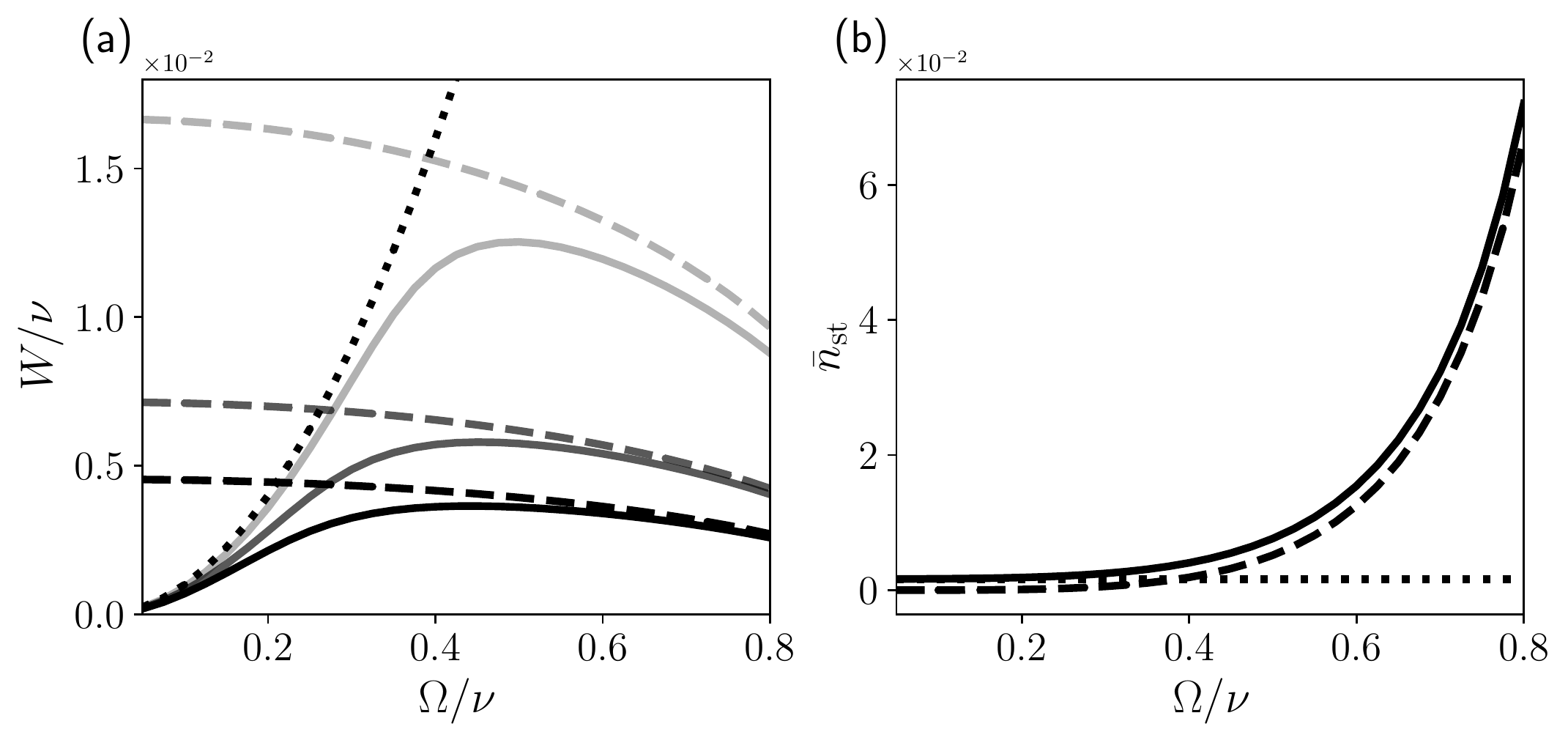} 
\par\end{centering}
\caption{(a) The cooling rate $W$ as a function of the laser coupling strength $\Omega$. The gray solid lines from darker to lighter correspond to the cooling rates fitted from the exact time evolution for $n_0=10,6,2$ respectively, while the corresponding dashed lines are our analytical predictions in the SSC regime with Eq.(\ref{eq:rate_ssc}). The black dotted line is the analytical prediction in the WSC regime with Eq.(\ref{eq:rate_wsc}). 
(b) The steady state average phonon occupation $\nss$ as a function of the laser coupling strength $\Omega$. 
The black solid line corresponds to the exact solutions of Eq.(\ref{eq:lindblad}). The dashed and dotted lines are analytical predictions from Eq.(\ref{eq:nss}) and Eq.(\ref{eq:nss_wsc}) respectively.
The other parameters used in these simulations are $\eta=0.1$, $\gamma=0.1\nu$. 
}
\label{fig:fig3} 
\end{figure}

Now we compare the analytical cooling rates and steady state average phonon occupations predicted in the SSC and WSC limits, with their corresponding values obtained from exponential fitting of the exact dynamics of $\nave(t)$. The results are shown in Fig.~\ref{fig:fig3}.
From Fig.~\ref{fig:fig3}(a), we can clearly see that the analytical cooling rate $\rate^{\WSC}$ from Eq.(\ref{eq:rate_wsc}) derived in the WSC limit agrees well with the exact results for small $\Omega$, while our $\rate^{\SSC}$ from Eq.(\ref{eq:rate_ssc}) derived in the SSC limit agrees well with the exact results for large $\Omega$. Moreover, Eq.(\ref{eq:rate_ssc}) correctly describes the dependence on $n_0$, while Eq.(\ref{eq:rate_wsc}) is completely independent of $n_0$. Additionally, the sideband coupling strength is also dependent on the motional state $\vert n\rangle$ and is proportional to $\eta\Omega \sqrt{n}$ (which can be seen from Eq.(\ref{eq:Hr2})), as a result for larger $n$ the sideband coupling is stronger and we are entering the SSC regime more quickly. Therefore for larger $n_0$, $\rate^{\SSC}$ agrees better with the exact results while $\rate^{\WSC}$ becomes worse ($\rate^{\WSC}$ deviates very quickly at $\Omega\approx 0.2\nu$ with the exact results for $n_0=2,6$). We can also see that for $n_0=10$ and $\Omega\approx 0.4\nu$, we can reach a maximal cooling rate of the order $10^{-2}\nu$. Defining the inverse trap frequency $\nu^{-1}$ as one cycle, this means that cooling could be achieved within hundreds of cycles for the standard sideband cooling in the SSC regime, while in comparison in the WSC regime thousands of cycles is usually required~\cite{roos1999quantum}.
From Fig.\ref{fig:fig3}(b), we can see that our analytical steady state average phonon occupation $\nss^{\SSC}$ in Eq.(\ref{eq:nss}) is slightly lower than the exact results, which is because that we have neglected the heatings from both the blue sideband transition and the recoil from the spontaneous decay. Nevertheless, our results are much more reasonable than the results predicted by $\nss^{\WSC}$ in Eq.(\ref{eq:nss_wsc}), since the latter is completely independent of $\Omega$ (For $\Omega < 0.4\nu$ the predictions from $\nss^{\WSC}$ is closer to the exact values). Therefore we can see that in the SSC limit, the blue sideband transition is indeed no longer the dominate heating source, which validates our approximation made in Eq.(\ref{eq:Hr}).


\begin{figure}[htbp]
\begin{centering}
\includegraphics[width=1\columnwidth]{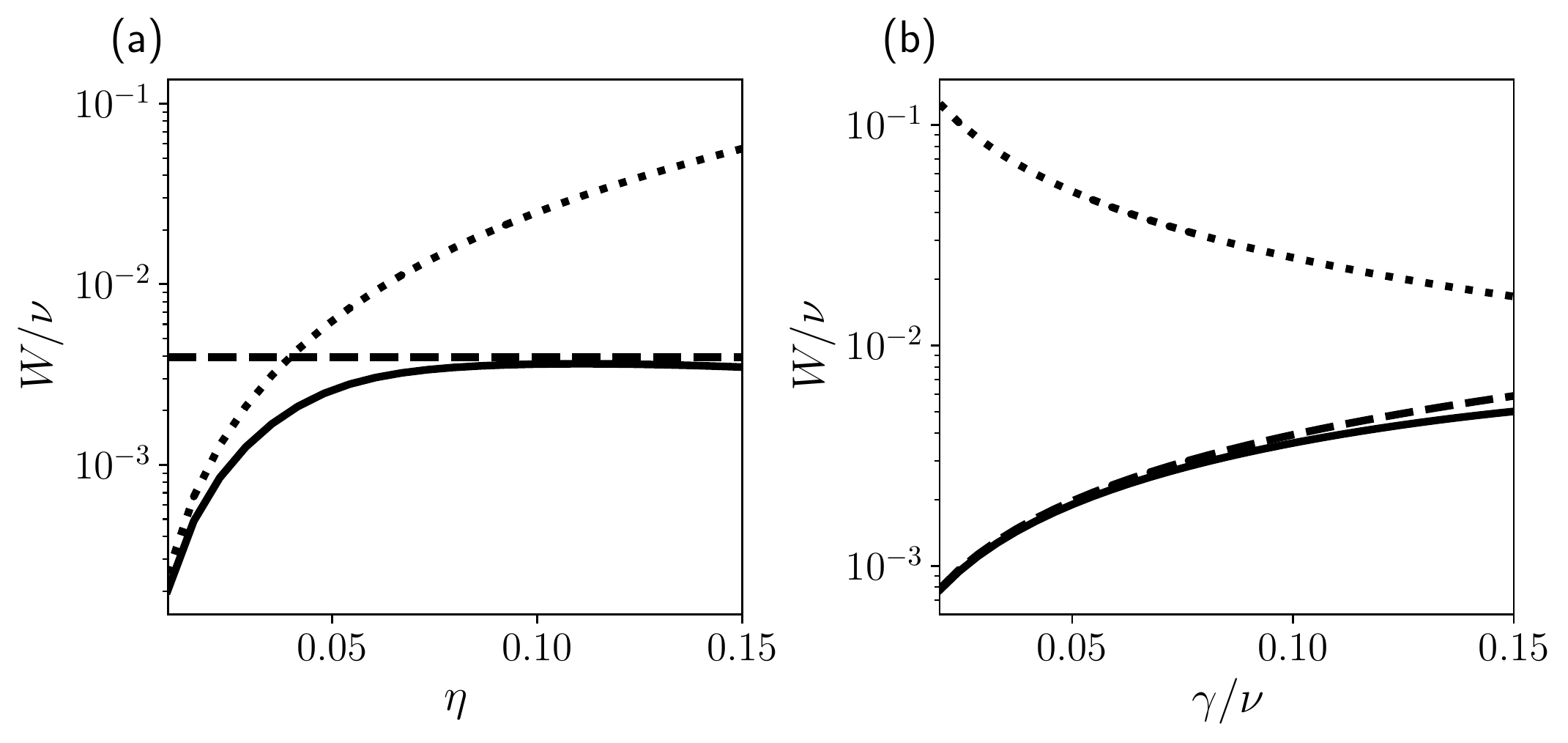} 
\par\end{centering}
\caption{The cooling rate $W$ as functions of (a) the Lamb-Dicke parameter $\eta$ and (b) the linewidth $\gamma$. The solid, dashed and dotted lines are results from the exact Lindblad equation, predicted by our strong sideband coupling theory and by the weak sideband coupling theory respectively.
We have used $\gamma=0.1\nu$ in (a) and $\eta=0.1$ in (b), and used $\Omega=0.5\nu$ and $n_0=10$ in both simulations.
}
\label{fig:fig4} 
\end{figure}

For completeness, we also study the dependence of the cooling rate on the Lamb-Dicke parameter $\eta$ and on the linewidth $\gamma$ since they also determine that we are in the SSC regime or not. The results are shown in Fig.~\ref{fig:fig4}.
From Fig.~\ref{fig:fig4}(a) we can see that $\rate^{\WSC}$ approximately agrees with the exact results for very small $\eta$, and starts to deviate significantly from the exact results for $\eta > 0.05$. In comparison, our $\rate^{\SSC}$ is independent of $\eta$ and becomes very accurate for $\eta > 0.1$, which is because that the sideband coupling strength increases with $\eta$. From Fig.~\ref{fig:fig4}(b), we can see the striking difference between $\rate^{\WSC}$ and $\rate^{\SSC}$: the former decreases with $\gamma$ and the latter increase with $\gamma$. Since we have chosen the parameters in the SSC regime, we can see that our $\rate^{\SSC}$ accurately predicts the cooling rates while $\rate^{\WSC}$ is completely off.


\section{Conclusion}\label{sec:summary}

In summary, we have considered the standard sideband cooling in the strong sideband coupling regime, and developed a theoretical framework to understand the cooling mechanism in this case. Under a modified red sideband resonance condition, we derive the analytical expressions for the cooling rate and the steady state average phonon occupation in the SSC regime. Our analytical predictions show that in the SSC regime a faster cooling rate proportional to the linewidth of the excited state can be achieved, and that the heating mechanism is strikingly different from that in the WSC regime. These results extend our previous results for dark-state cooling in the SSC regime to the more general situations with nonvanishing carrier transitions, which could also be straightforwardly applied to other cooling schemes with nonvanishing carrier transitions beyond the standard sideband cooling (for example the recent experimental work with the parallel EIT cooling method~\cite{zhang2022parallel}). 


\begin{acknowledgments}
The QuTip package is used for all the numerical simulations done in this work~\cite{qutip}.
C. G. acknowledges support from National Natural Science Foundation of China under Grants No.~11805279, No.~12074117, No.~61833010 and No.~12061131011.
\end{acknowledgments}

\bibliographystyle{apsrev4-1}
\bibliography{refs}

\end{document}